\documentclass[12pt,letterpaper]{article}
\usepackage{jheppub}
\usepackage{verbatim}
\usepackage{slashed}

\usepackage{graphicx}
\usepackage{subfigure}
\input{epsf}
\usepackage{epsfig}
\usepackage{epstopdf}
\usepackage{amsthm}

\usepackage{tikz}

\def\({\left(} \def\){\right)}
\def\[{\left[} \def\]{\right]}

\newcommand{\be}{\begin{equation}}
\newcommand{\ee}{\end{equation}}
\newcommand{\bea}{\begin{eqnarray}}
\newcommand{\eea}{\end{eqnarray}}
\newcommand{\ba}{\begin{eqnarray}}
\newcommand{\ea}{\end{eqnarray}}

\newcommand{\beq}{\begin{equation}}
\newcommand{\eeq}{\end{equation}}
\newcommand{\beqa}{\begin{eqnarray}}
\newcommand{\eeqa}{\end{eqnarray}}
\newcommand{\beqar}{\begin{eqnarray*}}
\newcommand{\eeqar}{\end{eqnarray*}}

\newcommand{\eg}{{\it e.g.,}\ }
\newcommand{\ie}{{\it i.e.}\ }

\newcommand{\tr}{{\rm tr}}

\title{A 4d non-BPS NS-NS microstate}

\author[a,b]{Soumangsu Chakraborty}
\author[a]{, Shaun D. Hampton}
\affiliation[a]{Institut de Physique Th\'eorique, Universit\'e Paris-Saclay, CNRS, CEA\\
Orme des Merisiers, 91191 Gif-sur-Yvette, France}
\affiliation[b]{Institute for Theoretical Physics, University of Amsterdam\\	
1090GL Amsterdam, The Netherlands}
\emailAdd{soumangsuchakraborty@gmail.com}
\emailAdd{shaun.hampton@ipht.fr}

\abstract{We construct a two-parameter four-dimensional non-BPS NS-NS smooth microstate solution that asymptotes to flat spacetime with a linear dilaton in type II superstring theory. From the microscopic point of view, the background is made out of a certain number of decoupled (\ie $g_s\to 0$) NS5 branes wrapping $T^3\times S^1\times S^1$ with fundamental strings wrapping non-contractable cycles of $S^1\times S^1$ with integer momentum modes along them. We show that perturbative worldsheet theory in this background is given by a null-gauged WZW model. We also show that the consistency of the  worldsheet theory imposes non-trivial constraints on the supergravity background.
}

\begin{document}
\maketitle
 
 \section{Introduction}

String theory provides a rich framework to address many questions about the microscopic nature of black holes \cite{Callan:1996dv,Das:1996wn,Das:1996ug,Maldacena:1996ix,David:1999ec,Maldacena:1999bp}. 
One of its most significant achievements is to count the number of BPS states of a certain D1-D5 system and show agreement with the Bekenstein-Hawking entropy of the corresponding black hole in the supergravity theory \cite{Strominger:1996sh,Maldacena:1999bp}.
This strongly suggests that there should exist a realization of black hole microstates within gravity. Since then much work has been done to understand this in a concrete way.

The main proponent in this direction is the fuzzball proposal \cite{Lunin:2001fv,Lunin:2001jy,Mathur:2005zp}, which says that generic black hole microstates are horizonless configurations of fundamental objects in string theory that have the same mass and charge as the corresponding black hole \footnote{See \cite{Bena:2022ldq,Bena:2022rna} and references therein for a recent review of the subject.}. The first smooth solutions, \cite{Lunin:2001fv,Lunin:2001jy} and \cite{Maldacena:2000dr}, which were constructed were two charge configurations. The geometries in \cite{Lunin:2001fv} were derived in the F1-P system in the $M_{4,1}\times S^1\times T^4$ setup with an F1 string with momentum P placed along $S^1$ where the string was given a nontrivial profile in $M_{4,1}$, a noncompact space. Through dualities this configuration can be mapped to the D1-D5 system with D1 branes wrapping $S^1$ and D5 branes wrapping $S^1\times T^4$ where the branes now have a nontrivial profile in $M_{4,1}$, \cite{Lunin:2001jy,Maldacena:2000dr}. 
A further refinement of the fuzzball proposal delineates i) generic and possibly non-geometric configurations, ii) singular configurations, and iii) smooth geometries. The third category has enjoyed a very fruitful exploration. These are called microstate geometries, configurations that are horizonless and smooth, admitting a nice realization in terms of supergravity. They have been studied in a variety of scenarios, \ie \cite{Giusto:2012yz,Mathur:2012tj,Bena:2015bea,Bena:2016ypk,Bena:2017xbt,Ceplak:2018pws,Heidmann:2019zws,Warner:2019jll,Govaerts:2023xxv}. 
Typically, three charge microstate geometries are constructed in the D1-D5-P system 
where the momentum charge, $P$, is a wave that travels along the common D1-D5 direction, $S^1$. In \cite{Mayerson:2020tcl,Ganchev:2021iwy} the authors used a consistent truncation from 10d down to 3d to describe a certain class of these microstate geometries. Similarly, from this 3d perspective, smooth, purely NS-NS microstate geometries were constructed in \cite{Ganchev:2022exf} and in \cite{Ceplak:2022pep} a set of NS-NS microstate geometries were derived in 10d. Though numerous examples have been constructed, the entropy coming from these configurations, however, is subleading in the powers of the charges \cite{Shigemori:2019orj,Mayerson:2020acj}. This is because the torus, $T^4$, on which the remaining directions of the D5 brane are wrapped, is largely a spectator in the full theory, containing no nontrivial dynamics. The reason being that, in the construction of such solutions, one typically smears over the torus directions resulting, upon reduction, in a theory described by 6d supergravity. The only effects along the $T^4$ are typically an overall breathing mode in which its volume oscillates according to the motion of the momentum wave along the common D1-D5 circle. In \cite{Bena:2022sge} the authors addressed a possibly singular limit of the three-charge solution in which a horizon might have formed. To resolve this, they constructed a geometry in the NS5-F1 frame with a nontrivial density profile along the common circle which carries the momentum $P$, and with spherical symmetry in the spacetime, $M_{4,1}$. This density profile, coming from a bound state of D0-D4 branes averts the formation of a macroscopic horizon by causing the common circle to pinch-off beforehand. Instead, the horizon area was shown to be of zero size and so near this point, the globally three-charge solution looked locally two-charge. This configuration belongs to the second class of solutions mentioned previously, namely those which are singular and even degenerate, corresponding to brane sources.

This work prompted the investigation of black hole microstates into the regime in which the $T^4$ was no longer treated democratically but allowed to have nontrivial behavior. Recent analysis \cite{Bena:2022wpl,Bena:2022fzf} suggests that in order to obtain the full entropy, at least parametrically in the charges, one must consider nontrivial dynamics along at least one direction of $T^4$. A step in this direction was made in \cite{Bena:2022wpl} where supersymmetry projectors were analyzed for three charge microstates in which one of the torus directions was taken to be nontrivial while maintaining spherical symmetry in the noncompact directions.  Two conditions were imposed: 1) that locally the object be half-BPS, a criterion that, so far, indicates that an object is horizonless, and 2) that the global symmetries are consistent with the presence of three global charges. These are statements coming purely from branes and supersymmetry and so should be valid irrespective of a geometrical description. This system of branes is argued to capture, parametrically, the three-charge entropy. In \cite{Bena:2022fzf} a similar procedure was performed but for a more general scenario where the underlying constituents had nontrivial behavior along the common circle, $S^1$, along one spacetime direction in $M_{4,1}$ and along one direction in $T^4$. See \cite{Li:2023jxb} for more recent work on locally enhanced supersymmetry in the context of three charge black hole microstates.

BPS configurations have been the focus of many studies of black hole microstates. This is because supersymmetry provides a reliable identification between gravitational physics and the dual holographic field theory. For works in this direction see \cite{Kanitscheider:2006zf,Kanitscheider:2007wq,Giusto:2015dfa,Giusto:2019qig,Ganchev:2021ewa,Martinec:2022okx}. In order to understand configurations whose properties are closer to the black holes we observe in nature, it is instructive to also study
non-BPS objects. From the perspective of black hole microstructure this has traditionally been a challenging task. However, progress has been toward this direction. The construction of a non-BPS microstate was originally carried out in \cite{Jejjala:2005yu} in which the configuration was highly rotating. 
More recently work has been done in computing non-BPS solutions in \cite{Ganchev:2021pgs}, which uses again a consistent truncation from the full 10d theory to a 3d. For work on the holography of non-BPS solutions see \cite{Ganchev:2021ewa}. A set of smooth nonextremal microstate geometries were constructed in \cite{Bombini:2017got}. A set of smooth non-BPS geometries were constructed in a variety of scenarios using non-trivial topological structures in \cite{Bena:2015drs, Bah:2021owp,Bah:2022yji,Bah:2022pdn}, with a smooth Schwarzschild-like geometry being constructed in \cite{Bah:2023ows}. 
While microstate geometries have yielded significant results in recent years one can consider what happens beyond the regime of supergravity. This is where string worldsheet methods play a significant role. Due to the difficulty of constructing the full worldsheet theory with Ramond-Ramond flux, worldsheet methods are most easily explored in NS-NS backgrounds. In \cite{Martinec:2017ztd} the authors constructed the string worldsheet solutions corresponding to two charge BPS microstates of a certain NS5-F1 configuration often termed as supertube geometries \cite{Mateos:2001qs,Emparan:2001ux}. Later, the construction of \cite{Martinec:2017ztd} was generalized to three-charge non-BPS microstates in \cite{Martinec:2018nco,Martinec:2019wzw,Martinec:2020gkv,Bufalini:2021ndn,Bufalini:2022wzu,Martinec:2022okx}. The worldsheet construction of the above-mentioned microstates are given by a certain class of null-cosets. 

In addition to the special class of null-cosets, discussed in the previous paragraph in the context of microstate geometries, worldsheet sigma models obtained by various gauging of $SL(2,\mathbb{R})$ WZW models have always played a very important role in understanding black hole physics \cite{Witten:1991yr,Dijkgraaf:1991ba,Giveon:2003ge,Chakraborty:2020yka}, solvable non-AdS holographic models \cite{Asrat:2017tzd,Chakraborty:2020yka,Chakraborty:2022dgm}, resolution of cosmological singularities \cite{Nappi:1992kv,Elitzur:2002vw,Elitzur:2002rt}  and condensed matter applications \cite{Goykhman:2013oja}. In this paper, we will take a similar approach as in \cite{Martinec:2017ztd} to construct a 
4d, non-BPS, NS-NS microstate solution that asymptotes to flat spacetime with a dilaton field that is linear in the radial direction. As in \cite{Martinec:2017ztd}, perturbative string theory in our microstate background is described by a certain null coset \eqref{cosetb}. We show that the consistency of the worldsheet theory ensures that the geometry is smooth (\ie no finite area horizon) and is labeled by four integers which are related to the background charges. Changing the background charges (which are allowed by the consistency of the worldsheet theory),  yields a different microstate solution. A closer inspection of the background charges reveals that the geometry is sourced by a certain number (greater than 1) of coincident NS5 branes wrapping $T^3\times S^1 \times S^1$ and F1 strings wrapping $S^1 \times S^1$ with winding and momentum charges along the two cycles. The smoothness of the geometry further imposes an algebraic constraint on the F1 winding and momentum charges. 

As discussed earlier, most NS5-F1 (or equivalently D1-D5 system in the S-dual frame) systems studied in the literature involve wrapping the fivebranes on $T^4\times S^1$ and F1 strings on $S^1$. In all such constructions, the $T^4$ is a mere spectator. From the supergravity point of view, exciting one or more moduli of the $T^4$ is technically challenging. One of the novel features of our worldsheet construction is to bypass this technical difficulty and put winding and momentum charges along one of the circles of $T^4$. In Appendix \ref{AppxA}, we generalize our construction even further where we excite all the moduli of $T^4$ and put momentum in the space transverse to the fivebranes (\ie separating the fivebranes rotating along a transverse circle and giving them some rotational angular momentum). One can think of these more general solutions as the  fully backreacted supergravity backgrounds discussed in \cite{Larsen:1999uk} with pure NS-NS flux.

The paper is organized as follows. In section \ref{sec2}, we construct the null gauged sigma model and read off the spacetime metric, the B-field, and the dilaton. In section \ref{sec3}, we take an algebraic approach to construct the string theory spectrum and derive the various worldsheet constraints to be imposed on the supergravity solution derived in section \ref{sec2}. In section \ref{sec4}, we compute the various background charges and impose the worldsheet constraints on the supergravity solution and investigate its consequences. In section \ref{sec5}, we interpret the supergravity solution as an RG flow and discuss its AdS decoupling limit. Finally, in section \ref{sec6},  we discuss our results and point out various directions for future research.

\section{Null gauged WZW model}\label{sec2}

In this section, we would like to study type II superstrings in the coset background 
\begin{equation}\label{cosetb}
\frac{SL(2,\mathbb{R})\times U(1)_x\times \mathbb{R}_t\times U(1)_y}{U(1)_L\times U(1)_R} \times S^3\times T^3~,
\end{equation}
where the gauged $U(1)_L\times U(1)_R$ is null with embedding
\begin{equation}
\begin{aligned}\label{gaugejjbar}
&\mathcal{J}=l_1 \mathcal{J}_3+l_2\mathcal{J}_x+l_3 \mathcal{J}_t+l_4 \mathcal{J}_y~,\\
&\bar{\mathcal{J}}=r_1 \bar{\mathcal{J}}_3+r_2\bar{\mathcal{J}}_x+r_3 \bar{\mathcal{J}}_t+r_4 \bar{\mathcal{J}}_y~,
\end{aligned}
\end{equation}
where $\mathcal{J}_3,\bar{\mathcal{J}}_3$ are the holomorphic and anti-holomorphic timelike currents of $SL(2,\mathbb{R})$, $\mathcal{J}_{x,t,y},\bar{\mathcal{J}}_{x,t,y}$ are the holomorphic and anti-holomorphic currents of $U(1)_x,\mathbb{R}_t, U(1)_y$ of the upstairs theory. 
Let $n_5$ be the level of $SL(2,\mathbb{R})$ supersymmetric WZW model. The coordinates $x,y$ that parametrize $U(1)_{x,y}$ respectively are compact with periodicities
\begin{equation}\label{periodicity}
x\sim x+2\pi R_x~, \ \ \ y\sim y+2\pi R_y~.
\end{equation}
 The currents
 \begin{equation}
     J_{3,x,t,y}\equiv i\mathcal{J}_{{3,x,t,y}}~, \ \ \ \bar{J}_{3,x,t,y}\equiv i\bar{\mathcal{J}}_{{3,x,t,y}}~,
 \end{equation}
are normalized such that they satisfy the following OPE: \footnote{Unless otherwise states we work in the convention $\alpha'=1$.}
\begin{equation}
\begin{aligned}\label{jjope}
&J_3(z)J_3(0)\sim\frac{-n_5/2}{z^2}~,\\
&J_x(z)J_x(0)\sim\frac{1/2}{z^2}~,\\
&J_t(z)J_t(0)\sim\frac{-1/2}{z^2}~,\\
&J_y(z)J_y(0)\sim\frac{1/2}{z^2}~,
\end{aligned}
\end{equation}
and similarly for the anti-holomorphic components $\bar{J}_{3,x,t,y}$. The detailed derivation of the gauged WZW model can be found in Appendix \ref{AppxA} with further generalizations.

Since the gauge currents $\mathcal{J},\bar{\mathcal{J}}$ are null, the $\mathcal{J}\mathcal{J}$ and  $\bar{\mathcal{J}}\bar{\mathcal{J}}$ OPE are regular. This implies
\begin{equation}
\begin{aligned}
&-n_5l_1^2+l_2^2-l_3^2+l_4^2=0~,\\
&-n_5r_1^2+r_2^2-r_3^2+r_4^2=0~.
\end{aligned}
\end{equation}
Without loss of generality, one can set $l_1=r_1=1$. This would imply
\begin{equation}
\begin{aligned}\label{mnullgc}
&l_2^2-l_3^2+l_4^2=n_5~,\\
&r_2^2-r_3^2+r_4^2=n_5~.
\end{aligned}
\end{equation}
The null gauge currents take the form 
\begin{equation}
\begin{aligned}
&\mathcal{J}= \mathcal{J}_3+l_2\mathcal{J}_x+l_3 \mathcal{J}_t+l_4 \mathcal{J}_y~,\\
&\bar{\mathcal{J}}= \bar{\mathcal{J}}_3+r_2\bar{\mathcal{J}}_x+r_3 \bar{\mathcal{J}}_t+r_4 \bar{\mathcal{J}}_y~.
\end{aligned}
\end{equation}

In order to compute the gauged sigma model, let's consider $g\in SL(2,\mathbb{R})\sim SU(1,1)$ to be parametrized as
\begin{equation}
g=e^{\frac{i}{2}(\tau-\sigma)\sigma_3}e^{\rho\sigma_1}e^{\frac{i}{2}(\tau+\sigma)}~.
\end{equation}
The gauged WZW action on 
\begin{equation}
\frac{SL(2,\mathbb{R})_{n_5}\times U(1)_x\times \mathbb{R}_t\times U(1)_y}{U(1)_L\times U(1)_R} 
\end{equation}
is given by \cite{Quella:2002fk}
\begin{equation}\label{gagact}
S=S[g]+S[x]+S[t]+S[y]+\frac{1}{2\pi}\int d^2z(A\bar{\mathcal{J}}+\bar{A}\mathcal{J}-MA\bar{A})~,
\end{equation}
where $A,\bar{A}$ are the gauge fields and 
\begin{equation}\label{M}
M=2(n_5\cosh2\rho-l_2r_2+l_3r_3-l_4r_4)~.
\end{equation}
The gauged action \eqref{gagact} is Gaussian in the gauge fields $A,\bar{A}$.  Integrating out the gauge fields one obtains
\begin{equation}\label{gagact1}
S=S[g]+S[x]+S[t]+S[y]+\frac{1}{2\pi}\int d^2z \frac{\mathcal{J}\bar{\mathcal{J}}}{M}~,
\end{equation}
along with the dilaton field 
\begin{equation}\label{dilaton}
\Phi=\Phi_0-\frac{1}{2}\log \Delta~,
\end{equation}
where 
\begin{equation}\label{delta}
\Delta=\frac{M}{2}~.
\end{equation}
The gauged action \eqref{gagact1} is invariant under null $U(1)_L\times U(1)_R$. In order to have a spacetime interpretation of the gauged sigma model one needs to fix the gauge. Thus, after fixing the gauge 
\begin{equation}
    \tau=\sigma=0~,
\end{equation}
one obtains
\begin{equation}
\begin{aligned}\label{gaugewzw}
S=&\frac{1}{2\pi} \int d^2z \left[-\left(1-\frac{2l_3r_3}{\Delta}\right) \partial t\bar{\partial} t +\left(1+\frac{2l_2r_2}{\Delta}\right) \partial x\bar{\partial} x+\left(1+\frac{2l_4r_4}{\Delta}\right) \partial y\bar{\partial} y+n_5\partial\rho\bar{\partial}\rho\right. \\
&\left. +\frac{2l_3r_2}{\Delta}\partial t\bar{\partial}x+\frac{2l_2r_3}{\Delta}\partial x\bar{\partial}t +\frac{2l_3r_4}{\Delta}\partial t\bar{\partial}y+\frac{2l_4r_3}{\Delta}\partial y\bar{\partial}t +\frac{2l_2r_4}{\Delta}\partial x\bar{\partial}y+\frac{2l_4r_2}{\Delta}\partial y\bar{\partial}x\right]~.
\end{aligned}
\end{equation}

Using standard worldsheet techniques, one can easily read off the metric, the B-field, and the dilaton (after lifting it to 10 dimensions and restoring $\alpha'=l_s^2=1$) as 
\begin{equation}
\begin{aligned}\label{sugrab}
\frac{ds^2}{\alpha'}&=-\left(1-\frac{2l_3r_3}{\Delta}\right) dt^2+n_5 d\rho^2+\left(1+\frac{2l_2r_2}{\Delta}\right) dx^2+\left(1+\frac{2l_4r_4}{\Delta}\right) dy^2\\
&+\frac{2(l_3r_2+l_2r_3)}{\Delta} dtdx+\frac{2(l_4r_3+l_3r_4)}{\Delta} dtdy+\frac{2(l_4r_2+l_2r_4)}{\Delta}dxdy+ds^2_{S^3}/\alpha'\\ 
&+ds^2_{T^3}/\alpha'~,\\
\frac{B}{\alpha'}=&\frac{l_3r_2-l_2r_3}{\Delta}dt\wedge dx+\frac{l_3r_4-l_4r_3}{\Delta} dt\wedge dy+\frac{l_2r_4-l_4r_2}{\Delta}dx\wedge dy+n_5 \cos^2\theta~ d\phi\wedge d\psi~,\\
e^{2\Phi}=&\frac{e^{2\Phi_0} }{\Delta}~,
\end{aligned}
\end{equation}
where $\Phi_0$ is  the dilaton background and 
\begin{equation}
    \Delta=n_5\cosh2\rho-l_2r_2+l_3r_3-l_4r_4~.
\end{equation}
The metric on $S^3$ is given by
\begin{equation}
ds^2_{s^3}/\alpha'=n_5(d\theta^2+\cos^2\theta d\psi^2 +\sin^2\theta d\phi^2)~,
\end{equation}
where $ds^2_{T^3}$ denotes the standard metric on $T^3$.

One can explicitly check that supergravity background \eqref{sugrab} satisfies the 10d type II supergravity equations of motion for all values of $l_{2,3,4},r_{2,3,4}$ subject to the null-gauge constraints \eqref{mnullgc}. In the next section, we will show the consistency of the worldsheet theory in the coset background \eqref{cosetb}, imposes further constraints on the gauge parameters  $l_{2,3,4},r_{2,3,4}$.

\section{Worldsheet constraints on supergravity}\label{sec3}

In this section, we will take an algebraic approach to construct the spectrum of physical operators of type II string theory in \eqref{G/H} with pure NS-NS flux and derive the constraints on the geometry imposed by the consistency of the worldsheet theory. To begin with, let's consider type II string theory in
\begin{equation}\label{G/H}
\frac{SL(2,\mathbb{R})_{k_{sl}}\times U(1)_x\times \mathbb{R}_t\times U(1)_y}{U(1)_L\times U(1)_R} \times SU(2)_{k_{su}}\times T^3~,
\end{equation}
where $k_{sl}$ and $k_{su}$ are respectively the WZW  (supersymmetric) levels of $SL(2,\mathbb{R})$ and $SU(2)$. Criticality of the worldsheet theory (\ie the total worldsheet central charge of the matter sector is $15$) requires 
\begin{equation}
k_{sl}=k_{su}\equiv n_5~.
\end{equation}
The null $U(1)_L\times U(1)_R$ gauge currents are generated  by
\begin{equation}
\begin{aligned}\label{brst}
&J= J_3+l_2J_x+l_3 J_t+l_4 J_y~,\\
&\bar{J}= \bar{J}_3+r_2\bar{J}_x+r_3 \bar{J}_t+r_4 \bar{J}_y~.
\end{aligned}
\end{equation}
The normalizations of the currents $J_{3,x,t,y}$ are such that they satisfy the OPE algebra \eqref{jjope}. Switching the chiralities one can equivalently obtain the OPEs of the anti-holomorphic currents $\bar{J}_{3,x,t,y}$. 
Regularity of the $JJ$ and $\bar{J}\bar{J}$ OPEs implies
\begin{equation}\label{nullgaugeconst}
l_2^2-l_3^2+l_4^2=r_2^2-r_3^2+r_4^2=n_5~.
\end{equation}
Here, as before, we set $l_1=r_1=1$ without loss of generality.

The vertex operators in the coset sigma model \eqref{G/H} in the $(-1,-1)$ picture are given by
\begin{equation}\label{vertop}
V(z,\bar{z})=e^{-\varphi}e^{-\bar{\varphi}}V^{j;w}_{m,\bar{m}}e^{i(P_Lx_L+iP_Rx_R)}e^{iEt}e^{i(Q_Ly_L+Q_Ry_R)}V_{\mathcal{M}}~,
\end{equation}
where $\varphi,\bar{\varphi}$ are the worldsheet superconformal fields coming from the bosonization of the $\beta\gamma$ and $\tilde{\beta}{\tilde\gamma}$ system, $V^{j;w}_{m,\bar{m}}$ is a spectrally flowed (with integer $w$) $SL(2,\mathbb{R})$ vertex operator from the discrete or continuous series representation, $e^{iEt},e^{i(P_Lx_L+iP_Rx_R)},e^{i(Q_Ly_L+Q_Ry_R)}$, are respectively the plane wave vertex operators in $\mathbb{R}_t,U(1)_x,U(1)_y$, 
$V_{\mathcal{M}}$ is a vertex operator of $SU(2)_{n_5}\times U(1)^3$ WZW model with dimensions $(\Delta^\mathcal{M}_L,\Delta^\mathcal{M}_R)$. The charges $P_{L,R},Q_{L,R}$ are given by
\begin{equation}
\begin{aligned}\label{PQ}
&P_{L,R}=\frac{n_x}{R_x}\pm w_x R_x~,\\
&Q_{L,R}=\frac{n_y}{R_y}\pm w_y R_y~,
\end{aligned}
\end{equation}
where $n_{x,y}$ and $w_{x,y}$  are integers and $E\in \mathbb{R}$.

Using the parafermionic decomposition \cite{Argurio:2000tb} of $SL(2,\mathbb{R})_{n_5}\sim SL(2,\mathbb{R})_{n_5}/U(1)\times U(1)_Y$, one can bosonize the timelike $SL(2,\mathbb{R})_{n_5}$ currents $J_3,\bar{J}_3$ as 
\begin{equation}\label{J3paraf}
    J_3=-\sqrt{\frac{n_5}{2}}\partial Y_L~,  \ \ \ \    \bar{J}_3=-\sqrt{\frac{n_5}{2}}\bar{\partial} Y_R~.
\end{equation}
where $Y=Y_L+Y_R$ is a free field normalized such that 
\begin{equation}\label{YYope}
    Y_L(z)Y_L(0)\sim -\log z~, \ \ \   Y_R(\bar{z})Y_R(0)\sim -\log \bar{z}~.
\end{equation}
The spectrally flowed $SL(2,\mathbb{R})_{n_5}$ vertex operator $V^{j;w}_{m,\bar{m}}$, under parafermionic decomposition, can be expressed as \cite{Argurio:2000tb}
\begin{equation}\label{vparaf}
V^{j;w}_{m,\bar{m}}= \Phi^j_{m,\bar{m}}e^{\sqrt{\frac{2}{n_5}}\left[\left(m+\frac{n_5}{2}w\right)Y_L+\left(\bar{m}+\frac{n_5}{2}w\right)Y_R\right]}~,
\end{equation}
where $ \Phi^j_{m,\bar{m}}$  is the parafermionic part of the $SL(2,\mathbb{R})$ vertex operator whose OPEs with $J_3,\bar{J}_3$ are regular. Using \eqref{J3paraf}, \eqref{YYope}, and \eqref{vparaf}, it is easy to show 
\begin{equation}\label{j3vope}
J_3(z)V^{j;w}_{m,\bar{m}}(0)\sim \frac{\left(m+\frac{n_5w}{2}\right)}{z} V^{j;w}_{m,\bar{m}}(0)~.
\end{equation}

Similarly, the OPE of the currents $J_{x,t,y}$ with the plane wave vertex operators are given by
\begin{equation}\label{jplanewope}
    J_x(z)e^{i(P_Lx_L+P_Rx_R)}(0)\sim \frac{P_L/2}{z}~, \ \ \ J_t(z)e^{iEt}(0)\sim \frac{-E/2}{z}~, \ \ \ J_y(z)e^{i(Q_Ly_L+Q_Ry_R)}(0)\sim \frac{Q_L/2}{z}~.
\end{equation}
Switching the chiralities one obtains the equivalent OPEs of the anti-holomorphic components of the currents and the vertex operators.

Physical vertex operators of the form \eqref{vertop} must be gauge invariant. This can be imposed by demanding that string states described by vertex operators \eqref{vertop} must be annihilated by the null gauge charges
\begin{equation}\label{qbrst}
    Q_B=\oint dz J~, \ \ \ \ \ \bar{Q}_B=\oint d\bar{z}\bar{J}~.
\end{equation}
An equivalent way of saying this would be to say that the OPEs of the null currents \eqref{brst} with the vertex operators \eqref{vertop} are regular. This imposes the null-gauge constraints
\begin{equation}
\begin{aligned}\label{nullgconst}
&\left(m+\frac{n_5w}{2}\right)+\frac{l_2P_L}{2}-\frac{l_3E}{2}+\frac{l_4Q_L}{2}=0~,\\
&\left(\bar{m}+\frac{n_5w}{2}\right)+\frac{r_2P_R}{2}-\frac{r_3E}{2}+\frac{r_4Q_R}{2}=0~.
\end{aligned}
\end{equation}
Note that a detailed analysis of the BRST quantization of the gauged WZW model \eqref{G/H} would give $cQ_{B},\bar{c}\bar{Q}_B$, \eqref{qbrst}, as the BRST charges that square to zero where $c,\bar{c}$ are the holomorphic and anti-holomorphic fields of the $bc$ and $\bar b\bar c$ system. Thus the null-gauge constraints, \eqref{nullgconst}, can also be thought of as the BRST constraints on the vertex operators. We would like to emphasize the fact that $cQ_{B},\bar{c}\bar{Q}_B$ are BRST charges of the gauged WZW model and not of the  worldsheet string theory. The worldsheet BRST charges are constructed in Appendix \ref{AppxB}.

We are interested in vertex operators \eqref{vertop} that live in the physical Hilbert space of string theory in \eqref{G/H}. This means, in addition to the null-gauge constraints \eqref{nullgconst}, the vertex operators must satisfy the Virasoro constraints
\begin{equation}\label{vircons}
\begin{aligned}
&-\frac{j(j+1)}{n_5}-mw-\frac{n_5w^2}{4}+\frac{P_L^2}{4}-\frac{E^2}{4}+\frac{Q_L^2}{4}+N_L+\Delta^{\mathcal{M}}_{L}=\frac{1}{2}~,\\
&-\frac{j(j+1)}{n_5}-\bar{m}w-\frac{n_5w^2}{4}+\frac{P_R^2}{4}-\frac{E^2}{4}+\frac{Q_R^2}{4}+N_R+\Delta^{\mathcal{M}}_{R}=\frac{1}{2}~,
\end{aligned}
\end{equation}
where $j$ parametrizes the quadratic Casimir of $SL(2,\mathbb{R})$ and $N_{L,R}$ are the oscillator levels in $AdS_3$.

Naively, one might think that vertex operators \eqref{vertop} satisfying the null gauge constraints \eqref{nullgconst} and the Virasoro constraints \eqref{vircons} constitute all states in the Hilbert space of the string theory. But, as we are going to argue next, there are still some residual gauge redundancies that need to be fixed. To realize  the residual gauge freedom, let's define the following fields:
\begin{equation}\label{hlr}
\begin{aligned}
&H_{L}=-i\sqrt{\frac{n_5}{2}} Y_L+l_2x_L+l_3t+l_4y_L~,\\
&H_{R}=-i\sqrt{\frac{n_5}{2} }Y_R+l_2x_R+l_3t+l_4y_R~.
\end{aligned}
\end{equation}
It's easy to check that the OPEs of $H_{L,R}$ with respect to the null gauge currents $J,\bar{J}$, \eqref{brst}, are regular. This means, a physical vertex operator $V$, \eqref{vertop}, and $V.e^{i \alpha (H_L+H_R)}$ are gauge equivalent $\forall \alpha \in \mathbb{R}$. This implies 
\begin{equation}
\begin{aligned}\label{chargeginv}
&(w,P_L,E,Q_L)\sim(w,P_L,E,Q_L)+(\alpha,\alpha l_2, \alpha l_3,\alpha l_4)~,\\
&(w,P_R,E,Q_R)\sim(w,P_R,E,Q_R)+(\alpha,\alpha r_2, \alpha r_3,\alpha r_4)~.
\end{aligned}
\end{equation}
The identifications \eqref{chargeginv} impose further constraints on $\alpha,l_{2,3,4},r_{2,3,4}$:
\begin{equation}
\begin{aligned}\label{lrPQ}
&\alpha \in \mathbb{Z}~, &  \alpha l_2=\tilde{P}_L~, & \ \   \alpha l_4=\tilde{Q}_L~,\\
& & \alpha r_2=\tilde{P}_R~, &\ \   \alpha r_4=\tilde{Q}_R~,\\
& & l_3=r_3~,&
\end{aligned}
\end{equation}
where 
\begin{equation}
\begin{aligned}\label{PQtilde}
&\tilde{P}_{L,R}=\frac{\tilde{n}_x}{R_x}\pm\tilde{w}_x R_x~,\\
&\tilde{Q}_{L,R}=\frac{\tilde{n}_y}{R_y}\pm\tilde{w}_y R_y~.
\end{aligned}
\end{equation}
where $\tilde{n}_{x,y},\tilde{w}_{x,y},\tilde{w}\in\mathbb{Z}$. Without loss of generality, one can set  $\alpha=1$.
Solving \eqref{chargeginv}--\eqref{PQtilde} and \eqref{nullgaugeconst}  one obtains \footnote{Note that the worldsheet constraints \eqref{wsconsts} on the gauge parameters $l_{2,3,4},r_{2,3,4}$
can also be obtained by directly analyzing the null gauge constraints \eqref{nullgconst}, the Virasoro constraints \eqref{vircons} and the periodicities of the vertex operator \eqref{vertop}. See \cite{Martinec:2018nco} for an elaborate
 discussion. }
\begin{equation}
\begin{aligned}\label{wsconsts}
&l_2=\frac{\tilde{n}_x}{R_x}+\tilde{w}_x R_x~, \ \ l_4=\frac{\tilde{n}_y}{R_y}+\tilde{w}_y R_y~, \ \  r_2=\frac{\tilde{n}_x}{R_x} -\tilde{w}_x R_x~, \ \  r_4=\frac{\tilde{n}_y}{R_y}-\tilde{w}_y R_y~,\\
&l=l_3=r_3=\sqrt{\tilde{P}_L^2+\tilde{Q}_L^2-n_5}=\sqrt{\tilde{P}_R^2+\tilde{Q}_R^2-n_5}\\
&  \ \ \ \ \ \ \ \ \ \ \ \ \ \ =\sqrt{\left(\frac{\tilde{n}_x^2}{R_x^2}+\frac{\tilde{n}_y^2}{R_y^2}+\tilde{w}_x^2R_x^2+\tilde{w}_y^2R_y^2\right)-n_5}~.
\end{aligned}
 \end{equation}
In \eqref{wsconsts} we have chosen the positive branch for $l_3,r_3$. The negative branch would simply correspond to $t\to-t$. 
The fact that $l_3=r_3$ imposes the following constraint
\begin{equation}\label{nullsurface}
\tilde{P}_L^2-\tilde{P}_R^2+\tilde{Q}_L^2-\tilde{Q}_R^2=0~,
\end{equation}
implying $\tilde{n}_{x,y},\tilde{w}_{x,y}$ lie on a null cone in $\Gamma_{2,2}$.
Substituting \eqref{PQtilde} into \eqref{nullsurface} one obtains
\begin{equation}\label{nwt}
\tilde{n}_x\tilde{w}_x+\tilde{n}_y\tilde{w}_y=0~.
\end{equation}
The reality of the gauge angle $l$ in \eqref{wsconsts} further imposes the constraint 
\begin{equation}\label{rxryconst}
    \frac{\tilde{n}_x^2}{R_x^2}+\frac{\tilde{n}_y^2}{R_y^2}+\tilde{w}_x^2R_x^2+\tilde{w}_y^2R_y^2\geq n_5~,
\end{equation}
for all integer values of $\tilde{n}_{x,y}$ and $\tilde{w}_{x,y}$  satisfying \eqref{nwt}. Thus, the allowed values of $R_{x,y}$ heavily depend on the choice of the integers $\tilde{n}_{x,y}$ and $\tilde{w}_{x,y}$.

\section{Supergravity background} \label{sec4}

In this section, we will study the supergravity background \eqref{sugrab}, calculate the conserved charges, and impose the worldsheet constraints  derived in the previous section. We would like to stress the fact that the solution \eqref{sugrab} satisfies the type II supergravity equations of  motion for all values of the gauge angles $l_{2,3,4},r_{2,3,4}$ subject to the constraint \eqref{nullgaugeconst}, but the consistency of the worldsheet theory imposes further constraints \eqref{wsconsts}--\eqref{rxryconst} on the gauge angles which are otherwise not manifest at the supergravity level.

To begin with, let's recall the supergravity background derived in section \ref{sec2}:
\begin{equation}
\begin{aligned}\label{sugrab1}
&\frac{ds^2}{\alpha'}=-\left(1-\frac{2l_3r_3}{\Delta}\right) dt^2 + n_5 d\rho^2+\left(1+\frac{2l_2r_2}{\Delta}\right) dx^2+\left(1+\frac{2l_4r_4}{\Delta}\right) dy^2\\
&\ \ \ \ \  +\frac{2(l_3r_2+l_2r_3)}{\Delta} dtdx+\frac{2(l_4r_3+l_3r_4)}{\Delta} dtdy+\frac{2(l_4r_2+l_2r_4)}{\Delta}dxdy+ds^2_{S^3}/\alpha'+ds^2_{T^3}/\alpha'~,\\
&\frac{B}{\alpha'}=\frac{l_3r_2-l_2r_3}{\Delta}dt\wedge dx+\frac{l_3r_4-l_4r_3}{\Delta} dt\wedge dy+\frac{l_2r_4-l_4r_2}{\Delta}dx\wedge dy+n_5 \cos^2\theta~ d\phi\wedge d\psi~,\\
&e^{2\Phi}=\frac{e^{2\Phi_0} }{\Delta}~,
\end{aligned}
\end{equation}
with
\begin{equation}
    \Delta=n_5\cosh2\rho-l_2r_2+l_3r_3-l_4r_4~.
\end{equation}
The background \eqref{sugrab1} has a boundary at 
$\rho\to \infty$.  Near the boundary, the background \eqref{sugrab1}  asymptotes to two-dimensional flat spacetime times two circles with radii $R_{x,y}$ (finite) and a dilaton that grows linear in $\rho$:
\begin{equation}
\begin{aligned}\label{ldasymp}
&\frac{ds^2}{\alpha'}=-dt^2+kd\rho^2+dx^2+dy^2~,\\
&B= 0~,\\
&\Phi=\tilde{\Phi}_0-\rho~.
\end{aligned}
\end{equation}
where $\tilde{\Phi}_0$ is a constant.

The IR limit of the solution \eqref{sugrab1} is obtained by taking the limit $R_y\to\infty$ keeping $R_x$ fixed \footnote{Since there is complete democracy between $R_x$ and $R_y$, one can also obtain an equivalent  IR limit by taking $R_x\to \infty$ keeping $R_y$ fixed. However, choosing $R_x$ or $R_y$ as the parameter to flow to the IR breaks the democracy in $x$ and $y$.}. This limit is a bit tricky and should be considered after imposing all the worldsheet constraints into the supergravity solution. We postpone this discussion to section \ref{sec5}.

In the discussion that follows, we will argue that the supergravity background \eqref{sugrab1} is a microstate solution and derive constraints on the solution for which the background geometry is smooth (absence of horizon).
The absence of a horizon in microstate geometries is a result of topological changes within the geometry.\footnote{Another way to understand this is the following: the near horizon region of a black hole geometry continues infinitely due to the presence of a horizon. For a microstate geometry, however, the near horizon region ends smoothly due to the presence of nontrivial microstructure at the would be horizon scale. For more discussion see \cite{Lunin:2001jy,Lunin:2001dt}. } When branes wrap compact directions, due to their tension, the tendency is for those directions to shrink. Placing momentum along the branes produces an opposing effect in which they want to expand. The balance between these two effects stabilizes the size of the compact directions, allowing for a horizon to form. For microstate geometries with the same global charges, however, the compact directions are stabilized in a certain region of the geometry but not all the way down to the horizon scale. Rather, the momentum dilutes in such a way as to allow the compact directions to shrink to zero precisely as one approaches the would-be horizon, allowing the space to end smoothly (up to an orbifold singularity at $\rho=0$).
Therefore, let us analyze our supergravity background and thus show that, under suitable constraints, it is a non-BPS, microstate geometry; a solution that is non-BPS, smooth, and horizonless. The analysis below is similar to the one that appears in \cite{Bufalini:2021ndn}. 
Let's begin by looking at the induced metric (in the string frame) on a constant $\rho=\rho_0$ surface at a fixed time and then consider the limit $\rho_0\to 0$. This is given by
\begin{equation}\label{indmet}
\lim_{\rho_0\to 0}\det \begin{pmatrix}
1+\frac{2l_2r_2}{\Delta} & \frac{l_4r_2+l_2r_4}{\Delta} \\
 \frac{l_4r_2+l_2r_4}{\Delta}& 1+\frac{2l_4r_4}{\Delta}
\end{pmatrix}=-\frac{n_5(l_3-r_3)^2}{(n_5-l_2r_2+l_3r_3-l_4r_4)^2}~.
\end{equation}
Note that this quantity is negative definite and requiring the absence of  a horizon imposes \footnote{One can also independently arrive at \eqref{l3r3} by demanding the absence of close timelike curves.}
\begin{equation}\label{l3r3}
l_3=r_3~.
\end{equation}
This is precisely the constraint one obtains from the consistency of the worldsheet theory \eqref{lrPQ}.
Let's define 
\begin{equation}
l=l_3=r_3~.
\end{equation}
The determinant of the 4d (excluding $S^3$ and $T^3$) part of the metric is given by
\begin{equation}
-\frac{n_5^3\sinh^22\rho}{\Delta^2}~,
\end{equation}
which also smoothly goes to zero as $\rho\to0$. One can also check that the Ricci scalar with $l_3=r_3$ is  smooth everywhere. 

Diagonalizing the induced metric at $\rho=\rho_0$ and at a fixed time and then taking $\rho_0\to 0$, one can easily show that one of the circles shrinks to zero size in the limit $l_3=r_3$ with a conical deficit. This is reflected in the fact that the determinant of the induced metric \eqref{indmet} vanishes at $\rho_0=0$. The size of the other circle, on the other hand, remains finite.

The supergravity background \eqref{sugrab1} has a $B$-field with components along $tx,ty,xy$, $\phi\psi$ directions. This means that the background is sourced by NS5-branes wrapping $T^3$ and the $x$ and $y$ circles and  F1-strings wrapping the $x$ and $y$ circles. The metric has non-zero $tx,ty$ components which further imply that the F1-strings have momentum modes along the $x$ and $y$ compact directions. In the following subsection, we will compute the conserved fundamental charges namely the NS5 and F1 charges and the ADM mass and angular momenta along $x$ and $y$.

\subsection{Background charges}

As is standard in any gauge theory, the charges are calculated from the integrals of the field strength over various surfaces. In the discussion that follows, we define the NS5 and F1 charges denoted respectively by $Q_5$ and $Q_1$,  such that they are topological charges that count respectively the number of NS5 branes and F1 strings wrapping the $x$-circle and the $y$-circle. 
The NS5-brane charge is given by 
\begin{equation}\label{ns5charge}
Q_5=\frac{1}{(2\pi)^2 l_s^2}\int_{S^3}H=n_5~,
\end{equation}
where, $H=dB$. Similarly, the F1-string charge is given by
\begin{equation}\label{f1charge}
Q_1=\frac{1}{(2\pi)^6l_s^6}\int_{\partial \mathcal{M}_8} e^{-2\Phi}\star_{10}H~,
\end{equation}
where $\partial \mathcal{M}_8$ is a 7-dimensional boundary of the 8-dimensional compact manifold at a fixed time and $\rho$. The 7-form $\tilde{H}=\star_{10}H$ has the following non-vanishing components:
\begin{equation}\label{Htilde}
\begin{aligned}
&\tilde{H}_{t\rho xyz_1z_2z_3}=\frac{2l_s^6n_5\sinh2\rho}{\Delta}~,\\
&\tilde{H}_{x\theta \psi\phi z_1z_2z_3}=\frac{l_s^6n_5l(l_4-r_4)\sin2\theta}{\Delta}~,\\
&\tilde{H}_{y\theta \psi\phi z_1z_2z_3}=\frac{l_s^6n_5l(l_2-r_2)\sin2\theta}{\Delta}~,
\end{aligned}
\end{equation}
where $z_{1,2,3}$ denote coordinates on $T^3$. The volume form on the 7-cycle at $\rho\to\infty$ can formally be expressed as
\begin{equation}
d\Omega_{7-\text{cycle}}=dx\wedge d\Omega_{S^3}\wedge d\Omega_{T^3}+dy\wedge d\Omega_{S^3}\wedge d\Omega_{T^3}~.
\end{equation}
Using this measure the total F1 winding charge can be split up into two components 
\begin{equation}\label{q1}
Q_1=|Q_1^x+Q_1^y|~,
\end{equation}
where
\begin{equation}
\begin{aligned}\label{q1xy}
&Q_1^x=\frac{1}{(2\pi)^6l_s^6}\int dy d\Omega_{S^3} d\Omega_{T^3} e^{-2\Phi}\tilde{H}_{y\theta \psi\phi z_1z_2z_3}=\frac{n_5l(l_2-r_2)R_yv}{e^{2\Phi_0}}~,\\
&Q_1^y=\frac{1}{(2\pi)^6l_s^6}\int dy d\Omega_{S^3} d\Omega_{T^3} e^{-2\Phi}\tilde{H}_{y\theta \psi\phi z_1z_2z_3}=\frac{n_5l(l_4-r_4)R_xv}{e^{2\Phi_0}}~,
\end{aligned}
\end{equation}
where $v$ is such that the dimension of $T^3$ is given by
\begin{equation}
V_{T^3}=(2\pi)^3l_s^4v~.
\end{equation}
The charges $Q_1^{x,y}$ can be thought of as the topological winding charges of the F1 strings along the $x$ and $y$ circles respectively.

In order to compute the ADM charges (mass and angular momenta), it is often useful to cast the string frame background metric \eqref{sugrab1} in the form 
\begin{equation}
\begin{aligned}\label{met1}
ds^2=g_{tt}dt^2+n_5 d\rho^2 +g_{ab}(dx^a+\mathcal{R}^a dt)(dx^b+\mathcal{R}^bdt)~,
\end{aligned}
\end{equation}
where $(a,b)$ run over $(x,y)$ and
\begin{equation}
\begin{aligned}\label{gR}
&g_{tt}=-\frac{n_5\cosh^2\rho}{l^2+n_5\cosh^2\rho}~, &    g_{ab}=\begin{pmatrix}
1+\frac{2l_2r_2}{\Delta} & \frac{l_4r_2+l_2r_4}{\Delta} \\
 \frac{l_4r_2+l_2r_4}{\Delta}& 1+\frac{2l_4r_4}{\Delta}
\end{pmatrix}~,\\
&\mathcal{R}^x=\frac{l(l_2+r_2)}{2(l^2+n_5\cosh^2\rho)}~, &\mathcal{R}^y=\frac{l(l_4+r_4)}{2(l^2+n_5\cosh^2\rho)}~.
\end{aligned}
\end{equation}
The geometry \eqref{met1} with \eqref{gR} and the B-field have nonzero $tx$ and $ty$ components that do not fall off sufficiently fast. This means that the supergravity solution  has finite angular charges, $L^{(G,B)}_{x,y}$. Using standard techniques from general relativity, one can read off the angular velocities as  
\begin{equation}
\begin{aligned}\label{omegaxy}
&\Omega^x=-\lim_{\rho\to 0}\mathcal{R}^x=-\frac{l(l_2+r_2)}{2(n_5+l^2)}~,\\
&\Omega^y=-\lim_{\rho\to 0}\mathcal{R}^y=-\frac{l(l_4+r_4)}{2(n_5+l^2)}~.
\end{aligned}
\end{equation}
The angular velocities \eqref{omegaxy} can further be decomposed into left and right movers \footnote{From the supergravity point of view, the four angular velocities $\Omega_{L,R}^{x,y}$ can be understood  as follows. Dimensional reduction along $S^1_{x,y}$ gives rise to four $U(1)$ gauge fields (two coming from the metric and two coming from the B-field). From the value of the gauge fields at $\rho=0$, one can work out the chemical potential conjugate to the four $U(1)$ charges using standard methods. From the higher dimensional point of view, the four chemical potentials can be interpreted as angular velocities $\Omega_{L,R}^{x,y}$. Similarly, the four $U(1)$ charges of the compactified theory can be realized as the four angular charges of the higher dimensional theory.}
\begin{equation}
\begin{aligned}\label{omegaLR}
&\Omega^x_L=-\frac{ll_2}{2(n_5+l^2)}~,  &\Omega^x_R=-\frac{lr_2}{2(n_5+l^2)}~,\\
&\Omega^y_L=-\frac{ll_4}{2(n_5+l^2)}~,  &\Omega^y_R=-\frac{lr_4}{2(n_5+l^2)}~.
\end{aligned}
\end{equation}

As stated earlier, the solution \eqref{met1} with \eqref{gR} is invariant under constant shifts in $t,x,y$. The corresponding conserved charges can be evaluated using the covariant phase space formalism \cite{Barnich:2001jy,Compere:2018aar,Chakraborty:2022dgm} which we calculate next. 

The conserved angular charges can be derived using the 1-forms
\begin{equation}\label{k1form}
    k_G^a=\mathcal{G}_{a\mu}dX^\mu~,  \ \ \ k_B^a=\mathcal{B}_{a\mu}dX^\mu~, \ \ \ a=x,y~,
\end{equation}
by the Komar formula \footnote{That the  conserved charges can be expressed as Komar integrals, follow directly from the covariant phase space formalism developed in \cite{Barnich:2001jy} for spacetimes with arbitrary asymptote.}
\begin{equation}\label{komar}
    L^{(G,B)}_{a}=\frac{1}{\kappa_0^2}V_{S^3}V_{T^3}l_s\int \star(dk_{(G,B)}^a)e^{-2\Phi}~,
\end{equation}
where  $\mathcal{G}_{\mu\nu},\mathcal{B}_{\mu\nu}$ are the components of the metric and the B-field in \eqref{sugrab1}, and $V_{S^3,T^3,x,y}$ are respectively the volumes of the $S^3,T^3$ and the $x$ and $y$ circles at the boundary given by
\begin{equation}
V_{S^3}=2\pi^2 n_5^{3/2}l_s^3~, \ \ \ \ V_{T^3}=(2\pi)^3l_s^3v~, \ \ \ \ V_x=2\pi R_x l_s~, \ \ \ \  V_y=2\pi R_y l_s~,
\end{equation}
and 
\begin{equation}
    \kappa=\kappa_0 e^{\Phi_0}=\sqrt{8\pi G_N}~,
\end{equation}
$G_N$ is the 10-dimensional Newton constant.
One can also define the left and the right moving components of the angular momenta as
\begin{equation}\label{JL}
    J_a^{L,R}=\frac{L_a^{(G)}\pm L_a^{(B)}}{2}~, \ \ \ a=x,y~.
\end{equation}
Using \eqref{komar} with \eqref{k1form} and  \eqref{JL} one can write
\begin{equation}
\begin{aligned}\label{JLR}
&J_x^L=-\frac{2 V_{S^3} V_{T^3}V_x V_y}{l_s^2\kappa_0^2}ll_2~, & &J_x^R=-\frac{2 V_{S^3} V_{T^3}V_x V_y}{l_s^2\kappa_0^2}lr_2~,\\
&J_y^L=-\frac{2 V_{S^3} V_{T^3}V_x V_y}{l_s^2\kappa_0^2}ll_4~, & &J_y^R=-\frac{2 V_{S^3} V_{T^3}V_x V_y}{l_s^2\kappa_0^2}lr_4~.
\end{aligned}
\end{equation}

Similarly one can use the analogous Komar integral for the energy/mass
\begin{equation}\label{komarm}
    E=-\frac{1}{\kappa_0^2}V_{S^3}V_{T^3}l_s\int \star(dk_{(G)}^t)e^{-2\Phi}~,
\end{equation}
where $dk_{(G)}^t=\mathcal{G}_{t\mu}dX^\mu$ which gives
\begin{equation}\label{adm}
    E=\frac{2 V_{S^3} V_{T^3}V_x V_y}{l_s^2\kappa_0^2}l^2~.
\end{equation}

Let us end this subsection with a quick consistency check of the various ADM charges \eqref{JLR},\eqref{adm}, and the angular potentials \eqref{omegaLR} derived above. 
Free energy of supergravity solution \eqref{sugrab1} is given by
\begin{equation}\label{free}
\mathcal{F}=E-TS-\Omega . J=E-TS-\Omega_L^xJ^L_x-\Omega_L^yJ^L_y-\Omega_R^xJ^R_x-\Omega_R^yJ^R_y~,
\end{equation}
where $S$ is the entropy of the system and $T$ is its temperature. It has been proved that the solution \eqref{sugrab1} with \eqref{l3r3} is smooth with no finite area horizon \ie $S=0$. This is reminiscent of the fact that the solution \eqref{sugrab1} with \eqref{l3r3} describes a microstate. Thus substituting  \eqref{omegaLR},\eqref{JLR},\eqref{adm},\eqref{JLR} and $S=0$ in \eqref{free} one obtains 
\begin{equation}
\mathcal{F}=E-\Omega_L^xJ^L_x-\Omega_L^yJ^L_y-\Omega_R^xJ^R_x-\Omega_R^yJ^R_y=0~.
\end{equation}
This can also be cross-checked by explicit computation of the Euclidean classical supergravity action. In fact, this observation agrees with the fact that in the semi-classical approximation (at leading order), the free energy of a supergravity solution that asymptotes to linear dilaton in the UV always vanishes \cite{Kutasov:2001uf,Giveon:2005jv,Chakraborty:2022dgm}.

\subsection{Summary of the supergravity solution}

Let us close this section with a brief summary of the full supergravity solution with all the worldsheet constraints imposed. 
As explained before the background is sourced by  $n_5$ NS5 branes and $Q_1$, \eqref{q1} F1 strings. This allows us to express the background dilaton $\Phi_0$ in terms of $Q_1$.  Using \eqref{q1} and \eqref{q1xy} one obtains 
\begin{equation}\label{phi0}
e^{2\Phi_0}=\frac{n_5vl}{Q_1}|(l_4-r_4)R_x+(l_2-r_2)R_y|=\frac{2n_5 lv R_xR_y|\tilde{w}_x+\tilde{w}_y|}{Q_1}~.
\end{equation}

Imposing the worldsheet constraints \eqref{wsconsts}--\eqref{rxryconst} in the supergravity solution \eqref{sugrab1} with \eqref{phi0}, one obtains
\begin{equation}\label{constsol}
\begin{aligned}
\frac{ds^2}{\alpha'}=& -\left(-1+\frac{2l^2}{\Sigma}\right)dt^2 + 
n_5d\rho^2+\left(1+\frac{2(\tilde{n}_x^2-R_x^4\tilde{w}_x^2)}{\Sigma R_x^2}\right)dx^2\\
&+\left(1+\frac{2(\tilde{n}_y^2-R_y^4\tilde{w}_y^2)}{\Sigma R_y^2}\right)dy^2+\frac{2\tilde{n}_xl}{\Sigma R_x}dtdx+\frac{2\tilde{n}_yl}{\Sigma R_y}dtdy\\
&+\frac{2}{\Sigma}\left(\frac{\tilde{n}_x\tilde{n_y}}{R_xR_y}-\tilde{w}_x\tilde{w}_yR_xR_y\right)dxdy+ds^2_{S^3}/\alpha'+ds^2_{T^3}/\alpha'~,\\
\frac{B}{\alpha'}=&-\frac{2\tilde{w}_xR_xl}{\Sigma}dt \wedge dx-\frac{2\tilde{w}_yR_yl}{\Sigma}dt \wedge dy+\frac{2}{\Sigma}\left(\frac{\tilde{n}_y\tilde{w}_xR_x}{R_y}-\frac{\tilde{n}_x\tilde{w}_yR_y}{R_x}\right) dx \wedge dy\\
&+n_5 \cos^2\theta~ d\phi\wedge d\psi~,\\
e^{2\Phi}=&\frac{2n_5R_xR_yvl|\tilde{w}_x+\tilde{w}_y|}{Q_1\Sigma}~,
\end{aligned}
\end{equation}
where
\begin{equation}
\begin{aligned}\label{lsigma}
l=&\sqrt{\left(\frac{\tilde{n}_x^2}{R_x^2}+\frac{\tilde{n}_y^2}{R_y^2}+\tilde{w}_x^2R_x^2+\tilde{w}_y^2R_y^2\right)-n_5}~,\\
\Sigma=&n_5\cosh2\rho+2\tilde{w}_x^2R_x^2 +2\tilde{w}_y^2R_y^2 -n_5~.
\end{aligned}
\end{equation}
with $\tilde{n}_{x,y},\tilde{w}_{x,y}\in \mathbb{Z}$ subject to the constraint
\begin{equation}\label{nwcons}
\tilde{n}_x\tilde{w}_x+\tilde{n}_y\tilde{w}_y=0~.
\end{equation}
Few comments about the supergravity solution \eqref{constsol}--\eqref{nwcons} are in order:
\begin{enumerate}

    \item{The supergravity solution \eqref{constsol}--\eqref{nwcons} is an NS-NS microstate solution of type II string theory. The geometry is smooth (up to an orbifold singularity at $\rho=0$) with no finite-size horizon. Perturbative string theory in this background is described by the null coset \eqref{cosetb}. The solution  \eqref{constsol}--\eqref{nwcons} is non-BPS because none of the spacetime supersymmetry generators are annihilated by the worldsheet BRST charge (See appendix \ref{AppxB} for a detailed discussion on this).
    
    There is another way to see that the solution is non-BPS. For 6d supergravity, BPS solutions take a particular form which makes manifest the supersymmetry. See \cite{Giusto:2013rxa} for more details. However, our solution after reducing to 6d (\eg reducing on either the $x$ or the $y$ circle) can not be rearranged into such a form thus supporting the claim that the geometry is in fact non-BPS.
    
    Further justification that the  solution \eqref{constsol}-\eqref{nwcons} is a microstate can be obtained by looking at the conserved charges. Our solution carries the NS5 charge, $n_5$, and F1 charges, $\tilde n_x,\tilde n_y$ (momentum) and $\tilde w_x, \tilde w_y$ (winding) under the constraint that $\tilde n_x\tilde w_x +\tilde n_y\tilde w_y=0$.  Changing any of these charges produces a different background seen by an observer at $\infty$. This is in contrast with a system that has both global charges and dipole charges. Dipole charges usually integrate to zero for an asymptotic observer. However, as we zoom in towards the would-be horizon, these dipole charges begin to distinguish and make manifest, locally, the microscopic structure or different microstates. }

\item{The JMaRT solution \cite{Jejjala:2005yu} is a non-BPS three-charge configuration. However, this state is very atypical from the point of view of black hole microstates because it is highly rotating along the $S^3$. In a similar manner, the microstate constructed in this paper is also atypical where, now, the momentum is carried along $S^1_x$ and $S^1_y$ instead of $S^3$. It is still interesting nonetheless, to derive a non-BPS supergravity background from worldsheet string theory, which is exact to all orders in $\alpha'$.}

\item{The background \eqref{constsol}--\eqref{nwcons} can be thought of a the fivebrane decoupling limit (\ie $g_s\to 0$) of a stack of $n_5$ NS5 branes wrapping $T^3\times S^1_x\times S^1_y$, with F1 strings  wrapping the $x$ and $y$-circles with integer winding numbers $\tilde{w}_{x,y}$ respectively and $\tilde{n}_{x,y}$ units of momenta along the $x$ and $y$-circles. Note that the integers $\tilde{w}_{x,y}$ and $\tilde{n}_{x,y}$ are not all independent. The smoothness of the geometry (\ie no horizon condition) imposes the constraint \eqref{nwcons} on the integers $\tilde{w}_{x,y}$ and $\tilde{n}_{x,y}$. Thus three out of the four integers $\tilde{w}_{x,y}$ and $\tilde{n}_{x,y}$ are independent. }

\end{enumerate}

\section{RG flow interpretation}\label{sec5}

As stated in section \ref{sec4}, the background \eqref{constsol} asymptotes to 2d flat spacetime times two spacelike circles with a dilaton field that is linear in the radial direction \eqref{ldasymp} and a vanishing B-field (The B-field is non-vanishing in the $S^3
$.).  The full background \eqref{constsol}, however, can be thought of as a two-parameter family of solutions parametrized by $R_x$ and $R_y$. In this section, we are going to argue that there exist,  one-parameter families of solutions embedded in the two-dimensional $R_x$, $R_y$ parameter space which can be interpreted as an RG flow from linear dilaton spacetime in the UV to some $AdS_3$ (locally) in the IR.    From the holographic point of view, the boundary field theory  is a certain Little String Theory with two dimensionless parameters $\lambda_{x,y}=\alpha'/R_{x,y}^2$ that flows to a CFT$_2$ in the IR. In the discussion that follows, we will identify two one-parameter lines of theories that flows to some $AdS_3$ at large distances.

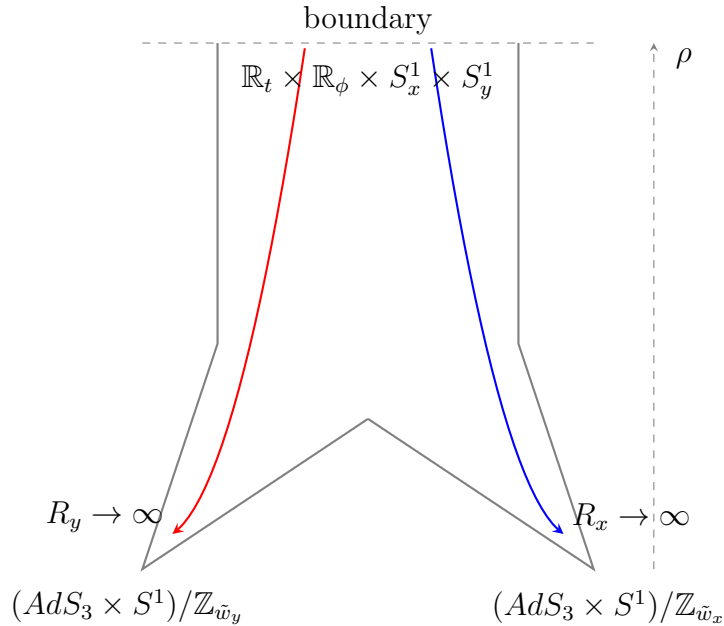
\begin{figure}
\begin{center}
\begin{tikzpicture}[scale=1, transform shape]
\draw[gray, thick] (0,0) -- (3,-2)--(2,1)--(2,5);
\draw[gray, thick] (0,0) -- (-3,-2)--(-2,1)--(-2,5);
\draw[gray, dashed] (-3,5)--(3,5);
\draw[stealth-,domain=.2:1.96, thick, red] plot ({\x-2.8}, {1.7*\x*\x-1.6});
\draw[-stealth,domain=-1.96:-.2, thick, blue] plot ({\x+2.8}, {1.7*\x*\x-1.6});
\draw (3.2,-2.5) node {$(AdS_3\times S^1)/\mathbb{Z}_{\tilde{w}_x}$}; 
\draw (-3.2,-2.5) node {$(AdS_3\times S^1)/\mathbb{Z}_{\tilde{w}_y}$};
\draw (3.5,-1.3) node {$R_x\to \infty$};
\draw (-3.5,-1.3) node {$R_y\to \infty$};
\draw (0,4.5) node {$\mathbb{R}_t\times \mathbb{R}_\phi \times S^1_x \times S^1_y$};
\draw[-stealth,gray, dashed] (3.8,-2) -- (3.8,5);
\draw (4.2,4.8) node {$\rho$}; 
\draw (0,5.3) node {boundary};
\end{tikzpicture}
  \end{center}
   \caption{Diagrammatic representation of the two-parameter family of the microstate solution \eqref{constsol}. The red curve denotes the flow obtained by varying $R_y$ keeping $R_x$ fixed. The IR of this flow is obtained when $R_y\to\infty$ where the spacetime is well approximated by $(AdS_3\times S^1)/\mathbb{Z}_{\tilde{w}_y}$. Similarly, the blue curve denotes the flow obtained by changing $R_x$ keeping $R_y$ fixed. The IR geometry here is given by $(AdS_3\times S^1)/\mathbb{Z}_{\tilde{w}_x}$.}
   \label{fig}
\end{figure}

Let's start by considering the one-parameter subspace in the $R_x$, $R_y$ parameter space obtained by varying $R_y$ for some fixed  $R_x$. The $AdS$ decoupling limit of such a one-parameter flow is obtained  by taking  $R_y\to \infty$ keeping $R_x$ fixed (see the red curve in figure \ref{fig}). In terms of the coordinates
\begin{equation}
\tilde{t}=\frac{t}{R_y}~, \ \ \ \tilde{y}=\frac{y}{R_y}~,
\end{equation}
which are well defined in the $R_y\to\infty$ limit, the supergravity background (suppressing $S^3\times T^3$) takes the form
\begin{equation}\label{IRads}
    \begin{aligned}
      &\frac{ds^2}{\alpha'}=n_5\left(-\frac{\cosh^2\rho}{\tilde{w}_y^2}d\tilde{t}^2+d\rho^2+\frac{\sinh^2\rho}{\tilde{w}_y^2}d\tilde{y}^2\right)+\left(dx+\frac{\tilde{n}_x}{\tilde{w}_yR_x}d\tilde{t}-\frac{\tilde{w}_xR_x}{\tilde{w}_y}d\tilde{y}\right)^2~,  \\
      &H= \frac{n_5\alpha'}{\tilde{w}_y^2}d\rho\wedge d\tilde{t}\wedge d\tilde{y}~,\\
   &   e^{2\Phi}=\frac{n_5 v R_x}{Q_1}\left|1+\frac{\tilde{w}_x}{\tilde{w}_y}\right|~.
    \end{aligned}
\end{equation}
Note that for $\tilde{w}_x=\tilde{n}_x=0$ and $\tilde{w}_y=1$, the above background is precisely  global-$AdS_3\times S^1$, but for generic $\tilde{w}_{x,y},\tilde{n}_x$, the background \eqref{IRads} can be identified as $(AdS_3\times S^1)/\mathbb{Z}_{\tilde{w}_y}$ with $\sqrt{n_5\alpha'}$ as the radius of $AdS$. Redefining the coordinate 
\begin{equation}\label{lgt}
x'= x+\frac{\tilde{n}_x}{\tilde{w}_yR_x}\tilde{t}-\frac{\tilde{w}_xR_x}{\tilde{w}_y}\tilde{y}~,
\end{equation}
with
\bea
x'\sim  x' + 2\pi R_x\bigg(1-{\tilde{w_x}\over\tilde{ w_y}}\bigg)~,
\eea
the metric takes the simpler form
\begin{equation}
      \frac{ds^2}{\alpha'}=n_5\left(-\frac{\cosh^2\rho}{\tilde{w}_y^2}d\tilde{t}^2+d\rho^2+\frac{\sinh^2\rho}{\tilde{w}_y^2}d\tilde{y}^2\right)+dx'^2~,
\end{equation}
This is known as a large gauge transformation because the effects are felt at the boundary. In the dual field theory, this is equivalent to a  spectral flow operation  of the $SL(2,\mathbb{C})$ invariant NS vacuum. It would be interesting to further explore the boundary interpretation of this background.\footnote{It's important to note that the large gauge transformation \eqref{lgt} is valid only in the AdS decoupling limit \eqref{IRads} and not in the full geometry \eqref{constsol}-\eqref{nwcons}.} This feature is similar to the JMaRT solution (See \cite{Bufalini:2021ndn} for details.).

Since the background \eqref{constsol} is democratic in $R_x$ and $R_y$, one can equivalently obtain a second line of theories by varying $R_x$ and keeping fix $R_y$. The $AdS$ decoupling limit is obtained by taking the limit $R_x\to\infty$ with $R_y$ kept fixed (the blue curve in figure \ref{fig}). The decoupled background in the IR, thus obtained,  is $(AdS_3\times S^1)/\mathbb{Z}_{\tilde{w}_x}$. 

It would be interesting to understand if there exists some other one-parameter family of solutions that would give rise to some (locally) $AdS_3$ in the IR. 

\section{Discussion}\label{sec6}

In this paper, we studied type II superstrings in the background \eqref{cosetb} in the presence of pure NS-NS flux. We showed that the corresponding supergravity background \eqref{constsol}-\eqref{nwcons} is a two-parameter (parametrized by $R_{x,y}$) family of smooth non-BPS microstate solution that satisfies the type II supergravity equations of motion. Microscopically one can think of  \eqref{constsol}-\eqref{nwcons} as the fivebrane decoupling limit of a stack of $n_5$ NS5 branes wrapping $T^3\times S^1_x \times S^1_y$ with F1 strings wrapping the non-contractable cycles of $S^1_x\times S^1_y$ with winding charges $\tilde{w}_{x,y}$ and $\tilde{n}_{x,y}$ unites of momenta respectively along the $x$ and $y$-circles. The background asymptotes to flat spacetime with a linear dilaton. The $AdS$ decoupling limit is obtained by sending $R_y \to \infty$ (or equivalently $R_x\to \infty$) keeping $R_x$ fixed (or equivalently $R_y$ fixed) where the background takes the form of some $\mathbb{Z}_w$ orbifold of $AdS_3\times S^1$.

The background derived in this paper is similar to the JMaRT solution, a highly rotating (atypical) three-charge non-BPS microstate. The main difference is that instead of carrying angular momentum along the $S^3$, our solution carries it along the $S_x^1$ and $S^1_y$ directions. 

The construction in this paper is intimately connected to single trace $T\bar{T}$ \cite{Giveon:2017nie,Chakraborty:2020swe,Chakraborty:2020yka,Chakraborty:2023mzc,Chakraborty:2023zdd} deformation of string theory in $AdS_3/\mathbb{Z}_N$ for some integer $N$. In particular if one sets $\tilde{n}_x$ and $\tilde{w}_x$ to zero then the coset describes  single trace $T\bar{T}$ deformation of the spacetime theory dual to string theory in global $AdS_3$ or its $\mathbb{Z}_N$ orbifolds.

Our construction paves the way for various follow-up problems. The first and foremost is to understand the structure of spatial entanglement using the Ryu-Takayanagi (RT) prescription. Microstate geometries are dual to coherent states of the spacetime theory.  The entanglement of a bipartite system in a coherent state has  certain specific features \cite{Giusto:2014aba}. It would be nice to revisit this issue from the behavior of the RT surface. Entanglement entropy in similar asymptotically linear dilaton backgrounds \cite{Chakraborty:2018kpr,Chakraborty:2020udr} exhibits a certain non-local features. It would be interesting to investigate such non-local effects in this setup as well.

One of the great advantages of having  a coset description is to construct operators of the spacetime theory. In fact, the correlation function of operators can easily be computed using techniques described in  \cite{Asrat:2017tzd,Chakraborty:2020yka,Chakraborty:2022dgm,Cui:2023jrb}. It would be interesting to set up a 4-point scattering problem in the background \eqref{constsol}-\eqref{nwcons} and understand its analytic properties.

For a better understanding of the microstate solution constructed in this paper, it is instructive to study solutions to Klein Gordon equation in this background. From the holographic point of view, this would allow us to compute correlation functions of operators of the spacetime theory. Introducing a perturbative polynomial potential would allow us to evaluate the Witten diagrams which would eventually lead to the computation of S-matrices in this background. It would also be interesting to probe the geometry by a D1 brane with its endpoints anchored at the conformal boundary.  As discussed in \cite{Chakraborty:2018aji}, in the context of a similar non-AdS setup, the classical effective action of a probe D1 brane captures various non-perturbative aspects of the spacetime theory (\eg various phases of the theory including confinement-deconfinement phase transitions). We hope to report on this in the future.

Most of the previous microstate geometries which have been constructed contain both Ramond-Ramond and NS-NS flux making a worldsheet formulation challenging. Recently, in \cite{Ganchev:2022exf,Ceplak:2022pep}, purely NS-NS microstate geometries were constructed. It would be interesting to explore the worldsheet formulation of these geometries. We plan to investigate this in future work.

\section*{Acknowledgements}

We thank I. Bena, J. de Boer, N. \v Ceplak, A. Giveon, A. Hashimoto, N. Kovensky, D. Kutasov, F. Larsen, E. Martinec, D. Turton, and N. Warner for useful discussions. We especially thank I. Bena for reading the manuscript and providing helpful insight and N. \v Ceplak for reading through the Introduction and providing helpful suggestions. The work of SC received funding under the Framework Program for Research and
“Horizon 2020” innovation under the Marie Skłodowska-Curie grant agreement n° 945298. This work was supported in part by the FACCTS Program at the University of Chicago. SC would like to thank APCTP, Pohang for hospitality during part of this work. SDH would like to thank the hospitality of the University of Warsaw, Faculty of Physics during the completion of this work.
The work of SDH is supported by ERC Grant 787320 - QBH Structure.

\appendix

\section{General null gauging}\label{AppxA}

In this appendix, we would like to study a general null gauging
\begin{equation}\label{geng}
   \frac{\mathbb{G}}{\mathbb{H}}\equiv \frac{SL(2,\mathbb{R})\times SU(2)\times U(1)^4\times \mathbb{R}_t\times U_y(1)}{(U(1)_L\times U(1)_R)_{null}}~,
\end{equation}
where we will refer to the WZW theory on $\mathbb{G}$ as the upstairs theory.
The four circles of $U(1)^4$ are parametrized by
$x_i, \ i=1,\cdots,4$ with periodicities $x_i\sim x_i+2\pi R_i$. As usual $y$ parametrizes $U(1)_y$ with radius $R_y$ and $t$ parametrizes $\mathbb{R}_t$. Let 
\begin{equation}
    G={\rm{diag}} \left[g,g',e^{i\sqrt{\frac{2}{n_5}}x_1},e^{i\sqrt{\frac{2}{n_5}}x_2},e^{i\sqrt{\frac{2}{n_5}}x_3},e^{i\sqrt{\frac{2}{n_5}}x_4},e^{-\sqrt{\frac{2}{n_5}}t},e^{i\sqrt{\frac{2}{n_5}}y}\right]~,
\end{equation}
with $g\in SL(2,\mathbb{R}), g'\in SU(2)$ be a diagonal block element of the upstairs theory \eqref{geng}. The $(U(1)_L\times U(1)_R)_{null}$ that we want to gauge are generated by $\exp{T_L}$ and $\exp{T_R}$ where $T_{L,R}$ are given by
\begin{equation}
    \begin{aligned}\label{symgen}
        &T_L={\rm{diag}}\left[il_1\sigma_3, il_2\sigma_3,i\sqrt{\frac{2}{n_5}}l_3,i\sqrt{\frac{2}{n_5}}l_4,i\sqrt{\frac{2}{n_5}}l_5,i\sqrt{\frac{2}{n_5}}l_6,\sqrt{\frac{2}{n_5}}l_7,i\sqrt{\frac{2}{n_5}}l_8\right]~,\\
      &  T_R={\rm{diag}}\left[ir_1\sigma_3, ir_2\sigma_3,i\sqrt{\frac{2}{n_5}}r_3,i\sqrt{\frac{2}{n_5}}r_4,i\sqrt{\frac{2}{n_5}}r_5,i\sqrt{\frac{2}{n_5}}r_6,\sqrt{\frac{2}{n_5}}r_7,i\sqrt{\frac{2}{n_5}}r_8\right]~.
    \end{aligned}
\end{equation}
In other words, we would like to gauge the symmetry 
\begin{equation}
    G\to e^{T_L} G e^{T_R}~.
\end{equation}
The generators of the symmetry \eqref{symgen} are null
\begin{equation}\label{nulltr}
    \tr(P.T_L^2)=\tr(P.T_R^2)=0~,
\end{equation}
where
\begin{equation}\label{P}
    P={\rm{diag}}\left[1,1,-1,-1,-1,-1,-1,-1,-1,-1\right]~,
\end{equation}
is the projection operator that keeps track of the signature of the Killing form \footnote{The projection operator is chosen such that the sign of the kinetic terms of the timelike fields are negative and the spacelike fields are positive.}.  Substituting \eqref{symgen} and \eqref{P} in \eqref{nulltr} one obtains,
\begin{equation}
    \begin{aligned}\label{lrnull}
        &-n_5l_1^2+n_5l_2^2+\sum_{i=3}^6l_i^2-l_7^2+l_8^2=0~,\\
         &-n_5r_1^2+n_5r_2^2+\sum_{i=3}^6r_i^2-r_7^2+r_8^2=0~.
    \end{aligned}
\end{equation}
Without loss of generality, we will set $l_1=l_2=1$.

The gauge currents are given by
\begin{equation}\label{Jgen}
    \begin{aligned}
        \mathcal{J}=n_5\tr (P.T_L.\partial G.G^{-1})~,\\
      \bar{\mathcal{J}}=n_5\tr (P.T_R.G^{-1}.\bar{\partial} G)~.
    \end{aligned}
\end{equation}

To compute the gauged sigma model let's choose the following parametrization for $g\in SL(2,\mathbb{R})$ and $g'\in SU(2)$
\begin{equation}\label{ggppar}
\begin{aligned}
   & g=e^{\frac{i}{2}(\tau-\sigma)\sigma_3}e^{\rho\sigma_1}e^{\frac{i}{2}(\tau+\sigma)\sigma_3}~,\\
   & g'=e^{\frac{i}{2}(\psi-\phi)\sigma_3}e^{i\theta\sigma_1}e^{\frac{i}{2}(\psi+\phi)\sigma_3}~.
   \end{aligned}
\end{equation}
The gauged sigma model is given by
\begin{equation}\label{gwzeact}
    S_g=S[G]+\frac{1}{2\pi}\int d^2z (A\bar{\mathcal{J}}+\bar{A}\mathcal{J}-M A\bar{A})~,
\end{equation}
where $A,\bar{A}$ are the gauge fields, $S[G]$ is the WZW action given by
\begin{equation}
    S[G]=\frac{n_5}{4\pi} \int d^2z ~\tr(PG^{-1}\partial GG^{-1}\bar{\partial}G)+\frac{in_5}{24\pi}\int_{\mathcal{B}}\tr(PG^{-1}dG\wedge G^{-1}dG\wedge G^{-1}dG)~,
\end{equation}
and 
\begin{equation}\label{mmm}
    M=2\left(n_5\cosh2\rho-n_5l_2r_2\cos2\theta-\sum_{i=2}^6l_ir_i+l_7r_7-l_8r_8\right)~.
\end{equation}
Integrating out the gauge fields in \eqref{gwzeact}, one obtains
\begin{equation}\label{genintact}
    S_g=S[G]+\frac{1}{2\pi}\int d^2z~ \frac{\mathcal{J}\bar{\mathcal{J}}}{M}~.
\end{equation}
Substituting \eqref{Jgen}, \eqref{gwzeact}, and \eqref{mmm} into \eqref{genintact} and using the parametrizations \eqref{ggppar} one obtains the gauged WZW action. After fixing gauge $\tau=\sigma=0$ and using standard worldsheet techniques one can read off the metric, B-field, and the dilaton as
\begin{equation}
    \begin{aligned}
  &  \frac{ds^2}{\alpha'}= -\left(1-\frac{2l^2}{\Delta^2}\right)dt^2 +n_5 d\rho^2 +\mathcal{G}_{ij}dX^idX^j+\tilde{\mathcal{G}}_{ij}dY^i dY^j+A_\psi d\psi+A_\phi d\phi+A_tdt~,\\
 &  \frac{ B}{\alpha'}=\frac{n_5l\cos^2\theta}{\Delta}(r_2-l_2)dt\wedge d\psi+\frac{n_5l\sin^2\theta}{\Delta}(r_2+l_2)dt\wedge d\phi\\ 
 &\ \ \ \ \ + dt\wedge B_t+d\phi\wedge B_\phi+d\psi\wedge B_\psi+\mathcal{B}d\phi\wedge d\psi+\tilde{\mathcal{B}}_{ij} dY^i\wedge dY^j~, \\
       & e^{2\Phi}=\frac{e^{2\Phi_0}}{\Delta}~,
    \end{aligned}
\end{equation}
where $l_7=r_7=l$ which follows from the smoothness of the geometry, $\Delta=M/2$, the vectors $X^i,Y^j$ are defined as
\begin{equation}
     X^i=\{\psi,\theta,\phi\}~,\ \ \ Y^i=\{x_1,x_2,x_3,x_4,y\}~,
\end{equation}
$\mathcal{G},\tilde{\mathcal{G}}$ are given by
\begin{equation}
    \mathcal{G}=\left(
\begin{array}{ccc}
 n_5 \cos ^2\theta \left(1+\frac{2n_5 l_2  r_2 \cos ^2\theta }{\Delta }\right) & 0 & 0 \\
 0 & n_5 & 0 \\
 0 & 0 & n_5 \sin ^2\theta  \left(1-\frac{2n_5 l_2  r_2 \sin ^2\theta }{\Delta }\right) \\
\end{array}
\right)~,
\end{equation}
\begin{equation}
    \tilde{\mathcal{G}}= \left(
\begin{array}{ccccc}
 1+\frac{2 l_3 r_3}{\Delta } & \frac{l_4 r_3+l_3 r_4}{\Delta } & \frac{l_5 r_3+l_3 r_5}{\Delta } & \frac{l_6 r_3+l_3 r_6}{\Delta } & \frac{l_8 r_3+l_3 r_8}{\Delta } \\
 \frac{l_4 r_3+l_3 r_4}{\Delta } & 1+\frac{2 l_4 r_4}{\Delta } & \frac{l_5 r_4+l_4 r_5}{\Delta } & \frac{l_6 r_4+l_4 r_6}{\Delta } & \frac{l_8 r_4+l_4 r_8}{\Delta } \\
 \frac{l_5 r_3+l_3 r_5}{\Delta } & \frac{l_5 r_4+l_4 r_5}{\Delta } & 1+\frac{2 l_5 r_5}{\Delta } & \frac{l_6 r_5+l_5 r_6}{\Delta } & \frac{l_8 r_5+l_5 r_8}{\Delta } \\
 \frac{l_6 r_3+l_3 r_6}{\Delta } & \frac{l_6 r_4+l_4 r_6}{\Delta } & \frac{l_6 r_5+l_5 r_6}{\Delta } & 1+\frac{2 l_6 r_6}{\Delta } & \frac{l_8 r_6+l_6 r_8}{\Delta } \\
 \frac{l_8 r_3+l_3 r_8}{\Delta } & \frac{l_8 r_4+l_4 r_8}{\Delta } & \frac{l_8 r_5+l_5 r_8}{\Delta } & \frac{l_8 r_6+l_6 r_8}{\Delta } &1+ \frac{2 l_8 r_8}{\Delta }\\
\end{array}
\right)~,
\end{equation}
$A_{\psi,\phi,t}$ are given by
\begin{equation}
\begin{aligned}
    A_\psi=&\frac{2n_5\cos^2\theta}{\Delta}\Big{[}l \left(l_2+r_2\right)dt+(l_3 r_2+l_2 r_3) dx_1+(l_4 r_2+l_2 r_4)dx_2+(l_5 r_2+l_2 r_5)dx_3\\
   & +(l_6 r_2+l_2 r_6)dx_4+(l_8 r_2 + l_2 r_8)dy\Big{]}~,\\
   A_\phi=&\frac{2n_5\sin^2\theta}{\Delta}\Big{[} l \left(r_2-l_2\right)dt+(l_3 r_2-l_2 r_3)dx_1+(l_4 r_2-l_2 r_4)dx_2+(l_5 r_2-l_2 r_5)dx_3\\
   &+(l_6 r_2-l_2 r_6)dx_4+(l_8 r_2-l_2 r_8)dy\Big{]}~,\\
   A_t=& \frac{2l}{\Delta}\Big{[}  \left( r_3+ l_3\right)dx_1+ \left( r_4+ l_4\right)dx_2+ \left( r_5+ l_5\right)dx_3+ \left( r_6+ l_6\right)dx_4+ \left( r_8+ l_8\right)dy\Big{]}~,
    \end{aligned}
\end{equation}
$\mathcal{B},\tilde{\mathcal{B}}$ are given by
\begin{equation}
    \mathcal{B}=\frac {n_ 5\cos^2 \theta} {\Delta}\left (\Delta - 
   2 l_ 2  r_ 2n_ 5\sin^2 \theta \right)~,
\end{equation}
\begin{equation}
    \tilde{\mathcal{B}}=
    \left(
\begin{array}{ccccc}
 0 & \frac{l_3 r_4-l_4 r_3}{\Delta } & \frac{l_3 r_5-l_5 r_3}{\Delta } & \frac{l_3 r_6-l_6 r_3}{\Delta } & \frac{l_3 r_8-l_8 r_3}{\Delta } \\
 0 & 0 & \frac{l_4 r_5-l_5 r_4}{\Delta } & \frac{l_4 r_6-l_6 r_4}{\Delta } & \frac{l_4 r_8-l_8 r_4}{\Delta } \\
 0 & 0 & 0 & \frac{l_5 r_6-l_6 r_5}{\Delta } & \frac{l_5 r_8-l_8 r_5}{\Delta } \\
 0 & 0 & 0 & 0 & \frac{l_6 r_8-l_8 r_6}{\Delta } \\
 0 & 0 & 0 & 0 & 0 \\
\end{array}
\right)~,
\end{equation}
and $B_{\psi,\phi,t}$ are given by
\begin{equation}
\begin{aligned}
    B_\psi=&\frac{n_5 \cos^2 \theta}{\Delta} \Big{[} \left(l_2 r_3-l_3 r_2\right)dx_1+ \left(l_2 r_4-l_4 r_2\right)dx_2+ \left(l_2 r_5-l_5 r_2\right)dx_3\\
    &+ \left(l_2 r_6-l_6 r_2\right)dx_4 + \left(l_2 r_8-l_8 r_2\right)dx_5\Big{]}~,\\
    B_\phi=&-\frac { n_ 5\sin^2\theta} {\Delta}\Big {[}\left (l_ 3 r_ 2 + 
      l_ 2 r_ 3 \right) dx_ 1 + \left (l_ 4 r_ 2 + 
      l_ 2 r_ 4 \right) dx_ 2 + \left (l_ 5 r_ 2 + 
      l_ 2 r_ 5 \right) dx_ 3 \\
      &+ \left (l_ 6 r_ 2 + 
      l_ 2 r_ 6 \right) dx_ 4 + \left (l_ 8 r_ 2 + 
      l_ 2 r_ 8 \right) dx_ 5\Big {]}~,\\
      B_t=&\frac {l} {\Delta}\Big{[} \left (r_ 3 - l_ 3 \right)dx_ 1 + 
 \left (r_ 4 - l_ 4 \right)dx_ 2 + \left (r_ 5 - l_ 5 \right)dx_ 3 +
  \left (r_ 6 - l_ 6 \right)dx_ 4 +\left (r_ 8 - l_ 6 \right) dx_ 5\Big{]}~.
    \end{aligned}
\end{equation}

\section{Spacetime supersymmetry}\label{AppxB}

The supergravity background \eqref{constsol}--\eqref{nwcons} obtained via the null coset is \eqref{cosetb} is non-BPS. Below we sketch the worldsheet argument to prove this statement.

The numerator theory (i.e. the theory before gauging) is $10+2$ dimensional. Such a theory has  $12$ spacetime fermions. Let $\psi_{sl}^{\pm,3}$ be the worldsheet fermions corresponding to $SL(2,\mathbb{R})$, $\psi_{x,t,y}$ be the worldsheet fermionic superpartners  of $x,t,y$ respectively, $\psi_{su}^{\pm,3}$ be the worldsheet fermions corresponding to $SU(2)$ and $\psi_{7,8,9}$ be the worldsheet fermionic superpartners of $T^3$.  Let's bosonize the fermions as follows
\begin{equation}
\begin{aligned}
&\psi_{sl}^+\psi_{sl}^-=i\sqrt{2}H_1~,  \ \ \  &\psi_{su}^+\psi_{su}^-=i\sqrt{2}H_2~,  \ \ \  &\psi_{sl}^3\psi_{su}^3=i\sqrt{2}H_3~, \\
&\psi_{t}\psi_{y}=i\sqrt{2}H_4~,  \ \ \  & \psi_{x}\psi_{7}=i\sqrt{2}H_5~,  \ \ \  &\psi_{8}\psi_{9}=i\sqrt{2}H_6~.
\end{aligned}
\end{equation}
The bosonized fields $H_{a}$ are normalised such that 
\begin{equation}
H_a(z)H_b(w)\sim-\delta_{ab}\log(z-w)~.
\end{equation}

The spin fields are given by
\begin{equation}
S_\varepsilon=e^{\frac{i}{\sqrt{2}}\sum_{a=1}^6\varepsilon_aH_a}~,
\end{equation}
where $\varepsilon_a=\pm$ are the fermion polarizations. The worldsheet has the usual superconformal $\beta,\gamma$ fields as well as the $\tilde{\beta},\tilde{\gamma}$ fields due to null gauging. The fields $\beta,\gamma,\tilde{\beta},\tilde{\gamma}$ are bosonized as
\begin{equation}
\begin{aligned}
&\beta \gamma=-\partial \varphi~, \ \ \ &\gamma =e^{\varphi}\eta~, \ \ \ & \beta=e^{-\varphi}\partial \eta~,\\
&\tilde{\beta} \tilde{\gamma}=-\partial \tilde{\varphi}~, \ \ \ &\tilde{\gamma} =e^{\tilde{\varphi}}\tilde{\eta}~, \ \ \ & \tilde{\beta}=e^{-\tilde{\varphi}}\partial \tilde{\eta}~.
\end{aligned}
\end{equation}

The spacetime supersymmetry operators in the $-1/2$ picture are given by
\begin{equation}
Q_\varepsilon=\oint dz e^{-\frac{1}{2}(\varphi-\tilde{\varphi})}S_\varepsilon~.
\end{equation}
Switching the worldsheet chiralities one can similarly define the anti-holomorphic spacetime supercharge $\bar{Q}_{\varepsilon}$.
The background is BPS or non-BPS depending on whether $Q_\varepsilon$ is annihilated by the BRST operator or not.
The BRST operators are given by
\begin{equation}
\begin{aligned}
&Q_{brst}=\oint \left(cT+\gamma G+\tilde{c}\mathcal{J}+\tilde{\gamma}\lambda+(bc,\tilde b\tilde c\text{ and }\beta\gamma, \tilde{\beta}\tilde{\gamma} \text{ terms})\right)~,\\
&\bar{Q}_{brst}=\oint \left(\bar{c}\bar{T}+\bar{\gamma}\bar{G}+\bar{\tilde{c}} \bar{\mathcal{J}}+\bar{\tilde{\gamma}}\bar{\lambda}+(\bar b\bar c,\bar{\tilde b}\bar{\tilde c}\text{ and }\bar{\beta}\bar{\gamma}, \bar{\tilde{\beta}}\bar{\tilde{\gamma}} \text{ terms})\right)~,
\end{aligned}
\end{equation}
where $\mathcal{J},\bar{\mathcal{J}}$ are the worldsheet null currents that we want to gauge, given by
\begin{equation}
\begin{aligned}
&\mathcal{J}=l_1 \mathcal{J}_3+l_2J_x+l_3 J_t+l_4 J_y~,\\
&\bar{\mathcal{J}}= r_1\bar{\mathcal{J}}_3+r_2\bar{J}_x+r_3 \bar{J}_t+r_4 \bar{J}_y~,
\end{aligned}
\end{equation}
with 
\begin{equation}
\begin{aligned}
&\mathcal{J}_3=J_3+\frac{2i}{k}\psi^1_{sl}\psi_{sl}^2~,\\
&\bar{\mathcal{J}}_3=\bar{J}_3+\frac{2i}{k}\bar{\psi}^1_{sl}\bar{\psi}_{sl}^2~.
\end{aligned}
\end{equation}
and $\lambda,\bar{\lambda}$ are the superpartners of the gauge currents $\mathcal{J},\bar{\mathcal{J}}$ given by
\begin{equation}
\begin{aligned}
&\lambda=l_1\psi_{sl}^3+l_2\psi_x+l_3\psi_t+l_4\psi_4~,\\
&\bar{\lambda}=r_1\bar{\psi}_{sl}^3+r_2\bar{\psi}_x+r_3\bar{\psi}_t+r_4\bar{\psi}_4~.
\end{aligned}
\end{equation}
Here $c,\bar{c}$ are the usual worldsheet fields of the $bc$ and $\bar b\bar c$ systems, and $\tilde{c},\bar{\tilde{c}}$ are the Faddeev–Popov fields that one needs to introduce due to null gauging.  
The $\tilde{c}\mathcal{J}$ and $\bar{\tilde{c}} \bar{\mathcal{J}}$ terms in $Q_{brst}$ and $\bar{Q}_{brst}$ acting on $Q_\varepsilon$ and $\bar{Q}_\varepsilon$ gives $l_1\varepsilon_1$ and $r_1\bar{\varepsilon}_1$ respectively. Since we assumed $l_1\neq0, r_1\neq 0$ to begin with, under no circumstances, the background can be BPS.

\newpage


\providecommand{\href}[2]{#2}\begingroup\raggedright\endgroup

\end{document}